\documentclass[
 reprint,
 amsmath,amssymb,
 aps,showpacs
]{appolb}
\usepackage{epsfig}
\usepackage{bm}
\usepackage{mathptmx}
\usepackage{amsmath}
\usepackage{subfigure}
\usepackage{wrapfig}
\usepackage{color}

\allowdisplaybreaks[1]

\begin{document}
\title{
\begin{picture}(0,0)(0,0)%
   \put(250,50){\makebox(0,0)[l]{\textnormal{\normalsize J-PARC-TH-0067}}}%
   \end{picture}%
Non-perturbative production rate of photons with a lattice quark propagator: effect of vertex correction%
\thanks{Presented at Critical Point and Onset of Deconfinement 2016, Wroc{\l}aw, Poland.}%
}
\author{Taekwang Kim
  \thanks{kim@kern.phys.sci.osaka-u.ac.jp},
  Masayuki Asakawa
\address{Department of Physics, Osaka University, Osaka 560-0043, Japan.}
\and
Masakiyo Kitazawa
\address{Department of Physics, Osaka University, Osaka 560-0043, Japan. \\
  J-PARC Branch, KEK Theory Center, Institute of Particle
  and Nuclear Studies, \\
  KEK, 203-1, Shirakata, Tokai, Ibaraki, 319-1106, Japan.}
}
\maketitle
\begin{abstract}
  We analyze the production rate of photons from the thermal medium
  above the deconfinement temperature
  with a quark propagator obtained from a lattice QCD numerical simulation.
  The photon-quark vertex is determined gauge-invariantly,
  so as to satisfy the Ward-Takahashi identity.
  The obtained photon production rate shows a suppression
  compared to perturbative results.
\end{abstract}
  
\section{Introduction}
\label{sec:Intro}

The photon production yield is an important experimental observable
in relativistic heavy ion collisions, 
because it serves as a direct signal from a hot medium.
Recently, interesting experimental results on 
the $p_{\rm T}$ spectra~\cite{Adare:2014fwh,Adam:2015lda}
and its anisotropic flow~\cite{Adare:2015lcd} are measured.
Theoretically, the production rate can be calculated perturbatively,
and sophisticated analyses on the leading order \cite{AMY0112} and
next-to-leading order \cite{Ghiglieri:2013gia} have been performed
based on the hard thermal loop~(HTL) resummed
perturbation theory.
It, however, is known that the hot medium near the critical temperature,
$T_{\rm c}$, is a strongly coupled system.
Therefore, non-perturbative analysis is more desirable to calculate the
production rate relevant for relativistic heavy ion collisions.

In Ref.~\cite{ours}, the analysis of the production rate of virtual photons, 
observed as dileptons in experiments, at zero momentum
has been 
performed using a quark propagator obtained on a lattice simulation.
The vertex function is constructed so as to satisfy the Ward-Takahashi
identity.
In the present study, we apply this analysis to the study of 
the real photon production rate.

\section{Formalism of Photon Production Rate}
\label{sec:Formalism1}

The photon production rate per unit time
per unit volume is related to the retarded photon self energy
$\Pi_{\mu\nu}^R(\omega,\bm{q})$ as 
\begin{align}
  \omega\frac{d{\rm N}_\gamma}{d^3qd^4x}
  = -\frac{2}{(2\pi)^3}\frac{1}{{\rm e}^{\beta\omega}-1}{\rm Im}
  \Pi_\mu^{R,\mu}(\omega,\bm{q}), \label{eq:ProductionRate}
\end{align}
with the inverse temperature $\beta=1/T$~\cite{GaleKapusta}.

The full photon self energy in Matsubara formalism is written as
\begin{align}
  \Pi_{\mu\nu}(i\omega_m,\bm{q})=-\sum_{\rm f}e_{\rm f}^2T\sum_n\int
  \frac{d^3p}{(2\pi)^3}{\rm Tr_CTr_D}
       [S(P)\gamma_{\mu}S(P+Q)\Gamma_{\nu}(P+Q,P)], \label{eq:Trace}
\end{align}
with the full quark propagator $S(P)$ and the 
full photon-quark vertex $\Gamma_{\nu}(P+Q,P)$.
For notational simplicity,
the color, flavor and Dirac indices of $S(P)$
are suppressed.
$\omega_m=2{\pi}Tm$ and $\nu_n=(2n+1){\pi}T$ with integers $m$ and $n$
represent the Matsubara frequencies for bosons and fermions, respectively.
$Q_\mu = (i\omega_m,\bm{q})$ and $P_\mu = (i\nu_n,\bm{p})$ are four
momenta of the photon and quarks, respectively.
$e_{\rm f}$ is the electric charge of a quark
and the index ``f" represents the quark flavor.
${\rm Tr_C}$ and ${\rm Tr_D}$ are the traces over the color and Dirac
indices, respectively.

In the present study, we use a quark propagator obtained on the lattice
as the full quark propagator in Eq.~(\ref{eq:Trace}).
On the lattice with a gauge fixing, the imaginary-time correlator
$S_{\mu\nu}(\tau,\bm{p})$ can be measured, which is related
to the spectral function $\rho_{\mu\nu}(\nu,{\bm p})$ as
\begin{equation}
S_{\mu\nu}\left(\tau,\bm{p}\right) = \int_{-\infty}^{\infty}d\nu\frac{\mbox{e}^{\left(1/2-\tau/\beta\right)\beta \nu}}{\mbox{e}^{\beta \nu/2}+\mbox{e}^{-\beta \nu/2}}\rho_{\mu\nu}\left(\nu,\bm{p}\right).
\label{eq:S(tau)-rho}
\end{equation}
We take the Landau gauge and
color indices are suppressed.

When chiral symmetry is restored, the spectral function
can be decomposed with the projection operators
$\Lambda_{\pm}\left(\bm{p}\right)=\left(1\pm\gamma_0\hat{\bm{p}}\cdot\bm{\gamma}\right)/2$ as
\begin{equation}
  \rho(\nu,\bm{p}) 
  = \rho_+(\nu,p) \Lambda_+(\bm{p}) \gamma_0 
  + \rho_-(\nu,p) \Lambda_-(\bm{p}) \gamma_0,
\end{equation}
where $
  \rho_\pm(\nu,p) \equiv 
  \mbox{Tr}_{\rm D}
  \left[\rho\left(\nu,\bm{p}\right)\gamma_0
    \Lambda_\pm\left(\bm{p}\right)\right]/2
  $
and $p = |{\bm p}|$.

In Ref.~\cite{Kaczmarek_etal}, the quark correlator
in the Landau gauge is evaluated
on the lattice with the quenched approximation,
and the quark spectral function is deduced with the two-pole ansatz,
\begin{equation}
  \rho_+(\nu,p)
  = Z_+(p) \delta\left(\nu-\nu_+(p)\right)
  + Z_-(p) \delta\left(\nu+\nu_-(p)\right),
  \label{eq:2pole}
\end{equation}
where $Z_\pm(p)$ and $\nu_\pm(p)$ are the residues and 
dispersions of two quasi-particle states of the normal and plasmino modes, 
respectively.
\begin{figure}
\begin{center}
\includegraphics[scale=0.7]{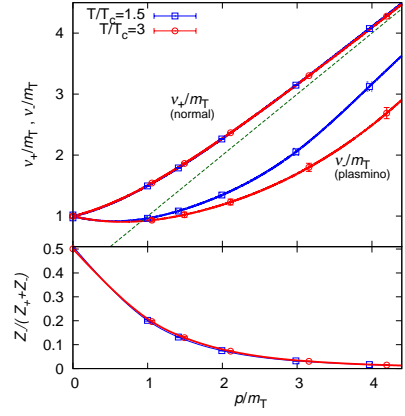}
\end{center}
  \caption{
    Open symbols show the momentum dependence of the parameters 
    $\nu_+(p)$, $\nu_-(p)$,
    and $Z_-(p)/(Z_+(p)+Z_-(p))$ obtained 
    on the lattice in Ref.~\cite{Kaczmarek_etal}.
    The solid lines represent their interpolation obtained 
    by the cubic spline method~\cite{ours}.
    The dashed line represents the light cone.
  }
\label{fig:one}
\end{figure}
In Fig.~\ref{fig:one}, we show the fitting result of each parameter
in Eq.~(\ref{eq:2pole}) for massless quarks as a function of $p$. 

Next, we construct the vertex function $\Gamma_\mu$.
Because of gauge invariance, this function should satisfy
the Ward-Takahashi identity,
\begin{align}
  Q^\mu\Gamma_\mu(P+Q,P) = S^{-1}(P+Q) - S^{-1}(P), \label{eq:WTI}
\end{align}
where $S^{-1}(P)$ is the inverse quark propagator with four momentum $P$.

In the present work, we use the following form of $\Gamma_\mu$,
\begin{align}
  &\Gamma_0(i\omega_m+i\nu_n,\bm{p}+\bm{q};i\nu_n,\bm{q})
  = \frac{1}{2i\omega_m} \left[S^{-1}(i\omega_m+i\nu_n,\bm{p}+\bm{q})
    - S^{-1}(i\nu_n,\bm{p}+\bm{q}) \right. \nonumber \\ 
    &\hspace{45mm}\left. + S^{-1}(i\omega_m+i\nu_n,\bm{q})
    - S^{-1}(i\nu_n,\bm{q}) \right], \label{eq:Gamma_0} \\
  &\Gamma_i(i\omega_m+i\nu_n,\bm{p}+\bm{q};i\nu_n,\bm{q})
  = \gamma_i-\frac{q_i}{2q^2} \left[S^{-1}(i\omega_m+i\nu_n,\bm{p}+\bm{q})
    + S^{-1}(i\nu_n,\bm{p}+\bm{q}) \right. \nonumber \\ 
    &\hspace{45mm}\left. - S^{-1}(i\omega_m+i\nu_n,\bm{q})
    - S^{-1}(i\nu_n,\bm{q}) \right]
  -\frac{q_i(\bm{q}\cdot\bm{\gamma})}{q^2} \label{eq:Gamma_i}
\end{align}
These vertex functions satisfy Eq.~(\ref{eq:WTI}).

With the lattice quark propagator and
the gauge invariant vertex Eqs.~(\ref{eq:Gamma_0}) and (\ref{eq:Gamma_i}),
the real photon production rate is evaluated as
\begin{align}
  &\omega\frac{d{\rm N}_\gamma}{d^3qd^4x}
  = -\frac{5\alpha}{6(2\pi)^3}\frac{1}{\omega}\frac{1}{{\rm e}^{\beta\omega}-1}
  \int_0^{\infty}dp_1\int_0^{\infty}dp_2 
  \sum_{s,t,\eta_1,\eta_2={\pm1}} Z_{\eta_1}(p_1)Z_{\eta_2}(p_2)
  \nonumber \\
  &\hspace{19mm}\times \biggl[ \biggl[ st\left[(sp_1+tp_2)^2-\omega^2\right] 
  \left[ 1-\frac{tp_2-sp_1}{\omega} \right]
  \nonumber \\
  &\hspace{19mm}\times 
    \left( 2+\frac{t\eta_2Z_{\bar{\eta}_2}(p_2)\bar{\nu}(p_2)}{s\eta_1\nu_{\eta_1}(p_1)+tV(p_2)}
    +\frac{s\eta_1Z_{\bar{\eta}_1}(p_1)\bar{\nu}(p_1)}{t\eta_2\nu_{\eta_2}(p_2)+sV(p_1)}
    \right)
    \nonumber \\
    &\hspace{19mm}
    -12p_1p_2+2st(sp_1+tp_2)^2
    -2st(p_1^2+p_2^2)+\frac{2st}{q^2}[(p_1^2-p_2^2)^2-q^4]
    \biggr]
    \nonumber \\
    &\hspace{19mm}\times
    \left[ f(s\eta_1\nu_{\eta_1}(p_1)) - f(t\eta_2\nu_{\eta_2}(p_2)) \right]
    \delta(\omega + s\eta_1\nu_{\eta_1}(p_1) - t\eta_2\nu_{\eta_2}(p_2))
  \nonumber \\
  &\hspace{19mm}
  - st\left[(sp_1+tp_2)^2-\omega^2\right] 
  \left[ 1-\frac{tp_2-sp_1}{\omega} \right]
  \frac{t\eta_2Z_{\bar{\eta}_2}(p_2)\bar{\nu}(p_2)}{s\eta_1\nu_{\eta_1}(p_1)+tV(p_2)}
  \nonumber \\
  &\hspace{19mm}
  \times
    \left[ f(-tV(p_2)) - f(t\eta_2\nu_{\eta_2}(p_2)) \right]
    \delta(\omega - tV(p_2) - t\eta_2\nu_{\eta_2}(p_2))
    \nonumber \\
    &\hspace{19mm}
    -st\left[(sp_1+tp_2)^2-\omega^2\right] \left[ 1-\frac{tp_2-sp_1}{\omega} \right]
    \frac{s\eta_1Z_{\bar{\eta}_1}(p_1)\bar{\nu}(p_1)}{t\eta_2\nu_{\eta_2}(p_2)+sV(p_1)}
    \nonumber \\
    &\hspace{19mm}\times
    \left[ f(s\eta_1\nu_{\eta_1}(p_1)) - f(-sV(p_1)) \right]
    \delta(\omega + s\eta_1\nu_{\eta_1}(p_1) + sV(p_1))
    \biggr]
  , \label{eq:VCrate}
\end{align}
where $\bar{\nu}(p_l)\equiv\nu_+(p_l)+\nu_-(p_l)$,
$V(p_l) \equiv Z_+(p_l)\nu_-(p_l)-Z_-(p_l)\nu_+(p_l)$
and $l$ takes $1$ or $2$.
We define $Z_{\pm1}(p_l) = Z_{\pm}(p_l)$,
$\nu_{\pm1}(p_l) = \nu_{\pm}(p_l)$
and ${\bar \eta}_l = -\eta_l$.
$f(p)$ is the Fermi distribution function.

The first term in Eq.~(\ref{eq:VCrate}) represents the
real photon productions via pair annihilation
and the Landau damping of quasi-quarks.
We note that the photon production with pair annihilation
can manifest itself in our formalism 
because of the modified dispersion relation of quasi-quarks.
On the other hand, the photon productions in
second and third terms cannot be interpreted as simple
reactions between quasi-quark excitations.
These anomalous photon production mechanism is found 
in our previous work on dilepton production rate~\cite{ours}.

\section{Numerical Results}
\label{sec:Results}

\begin{figure}[htbp]
\begin{center}
\subfigure[$T=1.5T_{\rm c}$]{
\includegraphics*[scale=0.665]{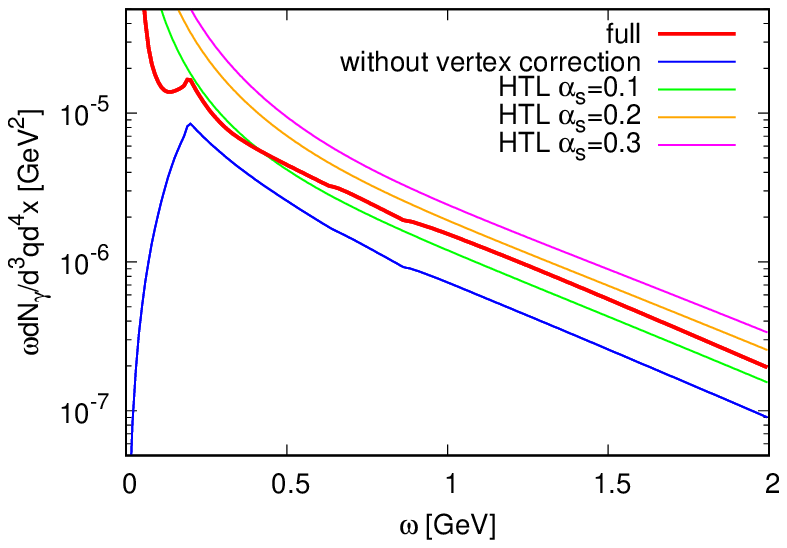}
\label{fig:two}}
\subfigure[$T=3T_{\rm c}$]{
\includegraphics*[scale=0.665]{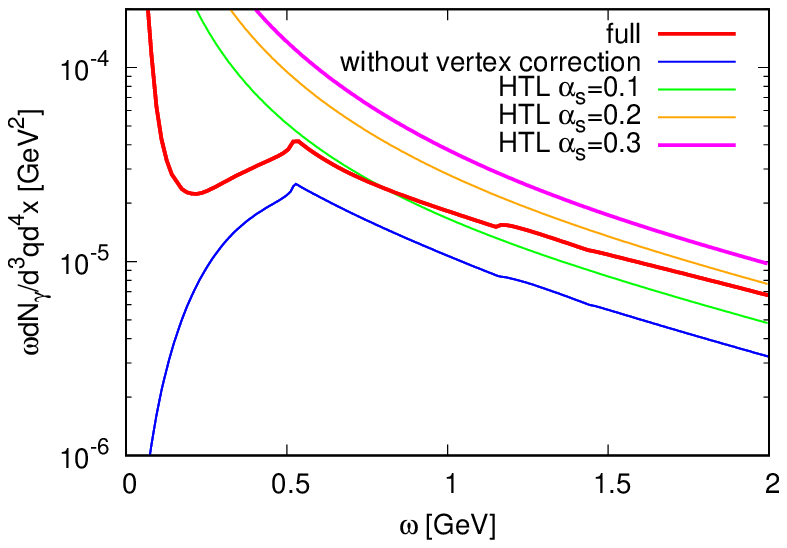}
\label{fig:three}}
\end{center}
\caption{
Photon production rates for $T=1.5T_{\rm c}$
and $3T_{\rm c}$.
The results without vertex correction are also plotted.
Thin lines represent the HTL results at the leading order.
}
\label{fig:results}
\end{figure}

Next, we show our numerical results of the photon production
rate obtained in the previous section.
In Fig.~\ref{fig:two}, we show
the energy dependence of the photon production rate
for $T = 1.5T_{\rm c}$.
In the figure, we also plot the result
calculated with bare vertex but the quark propagator obtained on the lattice.
Furthermore, the three thin lines represent the leading order result
in Ref.~\cite{AMY0112} for three values of strong coupling constant
$\alpha_s$.

In Fig.~\ref{fig:two}, one sees that our result is comparable to
perturbative ones.
It is interesting that the obtained result is similar to perturbative ones
although the production mechanisms are different.
In the perturbative calculation, the production rate is dominated by
bremsstrahlung and inelastic pair annihilation processes \cite{AMY0112}.
On the other hand, these processes are not directly included in our analysis.
Instead, the main contribution in our result comes from the Landau damping
and the pair annihilation processes of quasi-quark excitations.
Figure~\ref{fig:two} also shows that the production rate behaves
discontinuously at $\omega \simeq 0.63$~GeV.
The origin of this discontinuity is that the production rate is
given by a superposition of various reactions.
The pair annihilation process takes place for $\omega>0.63$~GeV
and the discontinuity corresponds to the threshold energy of this process.
Other structures can also be understood similarly.
In Fig.~\ref{fig:three}, the result for $T=3T_{\rm c}$ is shown,
which behaves similarly to the result for $T=1.5T_{\rm c}$.

\end{document}